\documentclass[aps,prd,onecolumn,groupedaddress,preprintnumbers,superscriptaddress,noshowpacs,nofootinbib]{revtex4-2}
\usepackage{amsmath,amssymb,latexsym}
\usepackage{graphicx}
\usepackage{color}
\usepackage{mathrsfs}

\usepackage{enumerate}
\usepackage[colorlinks=true,linkcolor=blue]{hyperref}

\begin{document}

\title{
Geometrical origin of the Kodama vector
}

\author{Shunichiro Kinoshita}
\email{kinoshita.shunichiro@nihon-u.ac.jp}
\affiliation{
Department of Physics, College of Humanities and Sciences,
Nihon University, Sakurajosui, Setagaya-ku, Tokyo 156-8550, Japan
}


\date{\today}

\begin{abstract}
 It has been known that warped-product spacetimes such as spherically
 symmetric ones admit the Kodama vector.
 This vector provides a locally conserved current made by 
 contraction of the Einstein tensor, even though there is no Killing vector. 
 In addition, a quasilocal mass, Birkhoff's
 theorem and various properties are closely related to the Kodama
 vector. 
 Recently, it is shown that the notion of the Kodama vector can be 
 extended to three-dimensional axisymmetric spacetimes even if the
 spacetimes are not warped product.
 This implies that warped product may not be a necessary condition for
 a spacetime to admit the Kodama vector.
 We show properties of the Kodama vector originate from the conformal
 Killing-Yano 2-form.
 In particular, the well-known spacetimes that admit the Kodama vector have
 a closed conformal Killing-Yano 2-form.
 Furthermore, we show the Kodama vector provides local conserved currents for
 each order of the Lovelock tensor as well as the Einstein tensor.
 \end{abstract}

\maketitle

 \section{Introduction and summary}
 
 The Kodama vector, which was at first found in four-dimensional
 spherically symmetric spacetimes~\cite{Kodama:1979vn}, provides a
 locally conserved current for the Einstein tensor even in spacetimes
 without Killing vectors such as dynamical spacetimes.
 Conventionally, the Kodama vector has been defined by $K^a = - \epsilon^{ab}\nabla_{b}r$, where $r$ denotes the areal radius and $\epsilon_{ab}$ denotes the two-dimensional volume form on time and radial space. 
 Since this vector $K^a$ satisfies $G^{ab} \nabla_a K_b = 0$ for the
 Einstein tensor $G_{ab}$, a current $J^a \equiv G^{ab}K_b$ is locally
 conserved, i.e., $\nabla_a J^a = 0$.
 If $K^a$ is timelike, this current can be interpreted as an appropriate energy
 current with assuming the Einstein equation and its associated charge
 yields the so-called Misner-Sharp quasilocal mass~\cite{Misner:1964je,Hayward:1994bu}.
 This notion has been generalized to higher dimensions straightforwardly.
 It is worth noting that spherical symmetry is not essential for a
 spacetime to admit the Kodama vector but warped-product with two-dimensional base space plays an important role.
 Moreover, it is known that the Kodama vector is closely related to Birkhoff's theorem (see~\cite{An:2017wti}, for example).
 This theorem states that all spherically symmetric solutions of the Einstein
 equation in vacuum must be static.
 It can be rephrased in terms of the Kodama vector as follows.
 The warped product spacetimes, including spherically symmetric spacetimes, admit the Kodama vector.
 If the spacetime is Einstein manifold, then the Kodama vector becomes the Killing vector. 

 Recently, it is shown that in three-dimensional axisymmetric spacetimes 
 even for nonwarped-product spacetimes such as rotating ones, the notion of the Kodama vector can be extended~\cite{Gundlach:2021six,Kinoshita:2021qsv}. 
 This vector can provide a local conserved current and quasilocal mass
 taking into account angular momentum, as in the cases of warped
 product spacetimes.
 This fact suggests that warped product does not seem to be necessary for a spacetime to admit the Kodama vector.

 In this paper we show properties of the Kodama vector geometrically
 originate from a conformal Killing-Yano (CKY) $2$-form.
 Various conserved currents and charges associated with (conformal)
 Killing tensors and (conformal) Killing-Yano forms have been reported
 in the literature~\cite{Jezierski:2002mn,Kastor:2004jk,Jezierski:2005cg,Lindstrom:2021qrk,Lindstrom:2021dpm,Gomez-Fayren:2023qly,Hull:2024xgo,Lindstrom:2022qjx,Ozkarsligil:2023avt}.
 What we emphasize here is that the Kodama vector is the so-called
 associated vector with a CKY $2$-form, while each subject has been discussed separately.
 In particular, all the well-known spacetimes admitting the Kodama vector have
 {\em closed} conformal Killing-Yano (CCKY) $2$-forms, which belong to a subclass of CKY $2$-forms.

 Furthermore, we show that the associated vector of the CKY $2$-form can
 yield conserved currents not only for the Einstein tensor but also for
 each order of the Lovelock tensor~\cite{Lanczos:1938sf,Lovelock:1971yv}.
 This means that the Kodama vector provides a locally conserved energy
 current in Lovelock gravity, which has been partially proved and
 conjectured for symmetric spacetimes such as spherically symmetric ones in~\cite{Maeda:2007uu,Maeda:2011ii}.
 (In warped-product spacetimes of a two-dimensional base and an Einstein space, the Kodama vector and the Misner-Sharp quasilocal mass were studied in Ref.~\cite{Ohashi:2015xaa}.)

 This paper is organized as follows.
 In Sec.~\ref{sec:CKY_Kodama} we present definitions and some basic properties
 of CKY $2$-forms.
 We reveal the relation between the Kodama vector and the associated vector of a CKY $2$-form. 
 In Sec.~\ref{sec:examples} we exhibit some explicit examples of the
 known Kodama vectors in terms of CKY $2$-forms.
 Thus, we demonstrate that the Kodama vectors arise from the CKY $2$-forms, indeed.

 \section{conformal Killing-Yano $2$-form and Kodama vector}
 \label{sec:CKY_Kodama}

 In this section we will show the associated vector of a conformal Killing-Yano $2$-form yields locally conserved currents contracted with each order of the Lovelock tensor, including the Einstein tensor. This implies that properties which the Kodama vectors should satisfy originate from conformal Killing-Yano $2$-forms. 

  \subsection{CKY $2$-form and conserved current for Einstein tensor}

 We consider that a $D$-dimensional spacetime with the metric $g_{ab}$
 admits a CKY $2$-form, $h_{ab}$.
 The conformal Killing-Yano $2$-form~\cite{Kashiwada:1968fva,tachibana1969conformal} (also, see~\cite{Frolov:2017kze} and references therein) satisfies 
 \begin{equation}
  \nabla_c h_{ab} = g_{ca} K_b - g_{cb} K_a + L_{abc},\quad
   K_a \equiv - \frac{1}{D-1}\nabla^b h_{ab} ,\quad
   L_{abc} \equiv \nabla_{[a}h_{bc]} ,
 \end{equation}
 where the vector field $K_a$ is the so-called associated vector of $h_{ab}$.
 If $L_{abc}=0$, $h_{ab}$ reduces to a CCKY $2$-form. 
 In this case, a Hodge dual of $h_{ab}$ yields a Killing-Yano $(D-2)$-form 
 $f_{a_1\cdots a_{D-2}}$, which satisfies 
 $\nabla_a f_{b_1 \cdots b_{D-2}}=\nabla_{[a}f_{b_1 \cdots b_{D-2}]}$.

 Covariant derivative of the associated vector $K_a$ is 
 \begin{equation}
  \begin{aligned}
   \nabla_a K_b &= - \frac{1}{D-1} \nabla_a \nabla^c h_{bc}
   = - \frac{1}{D-1} 
   \left(\nabla^c \nabla_a h_{bc} + R_a{}^c{}_b{}^d h_{dc}
   + R_a{}^c{}_c{}^d h_{bd}
   \right) \\
   &=  \frac{1}{D-1} \nabla_a K_b + \frac{1}{D-1} R_{acbd}h^{cd}
   + \frac{1}{D-1} R_a{}^c h_{bc}
   - \frac{1}{D-1} \nabla^c L_{abc} .
  \end{aligned}
 \end{equation}
 This can be rewritten as  
 \begin{equation}
   \begin{aligned}
    \nabla_a K_b 
    &= \frac{1}{2(D-2)} R_{abcd} h^{cd} 
    + \frac{1}{D-2} R_a{}^c h_{bc} - \frac{1}{D-2}\nabla^c L_{abc} ,
   \end{aligned}
   \label{eq:derivative_K}
 \end{equation}
 where we have used the first Bianchi identity 
 $R_{abcd} + R_{acdb} + R_{adbc} = 0$.

  It turns out that a symmetric part of Eq.~(\ref{eq:derivative_K}) is given by 
 \begin{equation}
  \nabla_{(a} K_{b)}
    = \frac{1}{D-2} R_{(a}{}^c h_{b)c} .
    \label{eq:symmetric_DK}
 \end{equation}
 The trace yields 
 \begin{equation}
  \nabla_a K^a = \frac{1}{D-2} R^{ac} h_{ac} = 0 ,
 \end{equation}
 implying that the vector field $K_a$ is divergence-free.
 For the Einstein tensor $G_{ab}$, we obtain 
 \begin{equation}
  G^{ab} \nabla_a K_b = \frac{1}{D-2} R^{ab} R_a{}^c h_{bc} = 0 .
 \end{equation}
 Thus, the associated vector $K^a$ for a conformal Killing-Yano $2$-form
 $h_{ab}$ 
 provides the same properties as Kodama vectors and $G_{ab}K^b$ becomes a locally conserved current.%
 \footnote{These properties have been pointed out in Refs.~\cite{Lindstrom:2021qrk,Lindstrom:2022qjx,Hull:2024xgo}, where $G_{ab}K^b$ is referred to as ``Einstein current.''}
 We note that if the spacetime is an Einstein space, i.e.,
 $R_{ab}=\lambda g_{ab}$, then Eq.~(\ref{eq:symmetric_DK}) leads to the 
 Killing equation $\nabla_a K_b + \nabla_b K_a = 0$~\cite{tachibana1969conformal}.
 This implies a version of Birkhoff's theorem that the Kodama vector becomes 
 a Killing vector in vacuum with a cosmological constant.
 In four dimensions, the relation between CKY $2$-form and Birkhoff's
 theorem was discussed~\cite{Ferrando:2015upa}.

 We can rewrite $G_{ab}K^b$ as 
 \begin{equation}
  \begin{aligned}
   G_{ab}K^b &= \frac{1}{2(D-3)} \nabla^b
   \left(R_{abcd}h^{cd} + 4 R_{[a}{}^c h_{b]c} + R h_{ab}\right) \\
   &= \nabla^b
   \left[\frac{1}{2(D-3)} W_{abcd}h^{cd} 
   + \frac{2}{D-2} R_{[a}{}^c h_{b]c}
   + \frac{D}{2(D-1)(D-2)} R h_{ab}
   \right] \\
   &= \frac{1}{2(D-2)} C_{abc} h^{bc}
   + \nabla^b
   \left[
   \frac{2}{D-2} R_{[a}{}^c h_{b]c}
   + \frac{D}{2(D-1)(D-2)} R h_{ab}
   \right] ,
  \end{aligned}
  \label{eq:Kodama_current}
 \end{equation}
 where $W_{abcd}$ denotes the Weyl curvature tensor and 
 the Cotton tensor $C_{abc}$ is defined as 
 \begin{equation}
  C_{abc} \equiv 2 \nabla_{[c}R_{b]a}
   - \frac{1}{D-1} g_{a[b} \nabla_{c]}R 
   = \frac{D-2}{D-3}\nabla^d W_{adbc} .
 \end{equation}
 Since $G_{ab}K^b$ is given by a divergence of $2$-form ``potential''
 in Eq.~(\ref{eq:Kodama_current}),
 we can explicitly see this current is locally conserved.
 It is worth noting that the expressions in the first and second lines
 of (\ref{eq:Kodama_current}) are 
 valid in $D>3$ dimensions, because both 
 $\mathcal{P}_{abcd} \equiv R_{abcd} - 2 R_{a[c}g_{d]b} + 2 R_{b[c}g_{d]a} + R g_{a[c}g_{d]b}$%
 \footnote{This rank-$4$ tensor is divergence-free and its indices have
 the same symmetries of the Riemann tensor, which can be also written as  
 $\delta_{abb_1b_2}^{cda_1a_2}R_{a_1a_2}{}^{b_1b_2}
 =4\mathcal{P}_{ab}{}^{cd}$ by using the generalized Kronecker $\delta$
 symbol.
 This type of tensor has been used in Ref.~\cite{Altas:2018pkl}, for example.
 }
 and $W_{abcd}$ are identically zero
 in three dimensions.
 However, that in the last line is valid even in $D=3$ dimensions.
 We note that $C_{abc}h^{bc}$ is a so-called Cotton current in
 Ref.~\cite{Lindstrom:2021dpm}.%
 \footnote{A conserved current for the Cotton tensor was discussed in
 Ref.~\cite{Perez:2010hk}, also.}

 In a specific case, if $h_{ab}$ is a Killing-Yano tensor, then the
 potential $2$-form field
 $R_{abcd} h^{cd} + 4 R_{[a}{}^c h_{b]c} + R h_{ab}$ itself can be conserved.
 This is referred to as the Yano current \cite{Kastor:2004jk}.
 It is equivalent to the fact that the associated vector for the
 Killing-Yano tensor will vanish in Eq.~(\ref{eq:Kodama_current}).

  \subsection{Generalization to Lovelock tensor}

 By using the fact that the Kodama vector is provided by a CKY $2$-form, we can prove the Kodama vector yields conserved currents for each order of the Lovelock tensor as well as the Einstein tensor.

 The $n$th order Lovelock tensor ($0<n<D/2$) in $D$
 dimensions~\cite{Lovelock:1971yv,Lanczos:1938sf} 
 (also, see \cite{Padmanabhan:2013xyr} and references therein) is given by 
 \begin{equation}
  G^{(n)}{}^a{}_{b} \equiv -
   \frac{1}{2^{n+1}} \delta^{aa_1 \cdots a_{2n}}_{bb_1\cdots b_{2n}}
   R_{a_1a_2}{}^{b_1b_2} \cdots R_{a_{2n-1}a_{2n}}{}^{b_{2n-1}b_{2n}} ,
 \end{equation}
 which reduces to the Einstein tensor for $n=1$.
 Note that symbol $\delta^{a_1 \cdots a_{k}}_{b_1\cdots b_{k}}$ is the generalized Kronecker $\delta$ symbol, defined by 
 \begin{equation}
  \delta^{a_1 \cdots a_{k}}_{b_1\cdots b_{k}}
   = k! g^{a_1}_{[b_1}\cdots g^{a_k}_{b_k]}
   = - \frac{1}{(D-k)!}
   \epsilon^{a_1 \cdots a_k c_{k+1} \cdots c_{D}} 
   \epsilon_{b_1 \cdots b_k c_{k+1} \cdots c_{D}} ,
 \end{equation}
 where $\epsilon_{a_1\cdots a_D}$ denotes the totally antisymmetric
 $D$-dimensional volume form.

We introduce the following $2$-form field consisting of a CKY $2$-form $h_{ab}$ and $n$ powers of the Riemann tensors: 
\begin{equation}
 F^{(n)}{}_{ab} \equiv \delta^{cda_1\cdots a_{2n}}_{abb_1\cdots b_{2n}}
  h_{cd}R_{a_1a_2}{}^{b_1b_2}\cdots R_{a_{2n-1}a_{2n}}{}^{b_{2n-1}b_{2n}} .
\label{eq:potential_two-form}
\end{equation}
It turns out that 
\begin{equation}
 \begin{aligned}
  \nabla^b F^{(n)}{}_{ab} &= \delta^{cda_1\cdots a_{2n}}_{abb_1\cdots b_{2n}}
  \nabla^b h_{cd}R_{a_1a_2}{}^{b_1b_2}\cdots R_{a_{2n-1}a_{2n}}{}^{b_{2n-1}b_{2n}} \\
  & \quad + \delta^{cda_1\cdots a_{2n}}_{abb_1\cdots b_{2n}}h_{cd}
  \sum_{k=1}^n
  R_{a_1a_2}{}^{b_1b_2}\cdots 
  \nabla^b R_{a_{2k-1}a_{2k}}{}^{b_{2k-1}b_{2k}}
  \cdots R_{a_{2n-1}a_{2n}}{}^{b_{2n-1}b_{2n}} \\
  &= \delta^{cda_1\cdots a_{2n}}_{abb_1\cdots b_{2n}}
  (g^b{}_c K_d - g^b{}_d K_c + L^b{}_{cd})
  R_{a_1a_2}{}^{b_1b_2}\cdots R_{a_{2n-1}a_{2n}}{}^{b_{2n-1}b_{2n}} \\
  &= - 2 (D-2n-1)\delta^{da_1\cdots a_{2n}}_{ab_1\cdots b_{2n}}
  K_d 
  R_{a_1a_2}{}^{b_1b_2}\cdots R_{a_{2n-1}a_{2n}}{}^{b_{2n-1}b_{2n}} \\
  &= 2^{n+2}(D-2n-1)G^{(n)}{}_{ad} K^d .
 \end{aligned}
\end{equation}
The second equality follows from the second Bianchi identity, 
$\nabla_{[a}R_{bc]de} = 0$, and the third equality does from the first
Bianchi identity.
Since $F_{ab}$ is antisymmetric, $G^{(n)}{}_{ad} K^d$ is divergence-free.
Hence, we have also a local conserved current for the $n$th order Lovelock tensor as  
\begin{equation}
J^{(n)}{}^a \equiv G^{(n)}{}^a{}_b K^b
 = \frac{1}{2^{n+2}(D-2n-1)} \nabla_b F^{(n)ab} .
\end{equation}
Note that, for $n=1$, the previous result for the Einstein tensor is obviously reproduced.
On arbitrary spacelike hypersurfaces $\Sigma$ with a common boundary
$\partial\Sigma$, by using the Stokes theorem, we have a conserved charge
written in the boundary integral. 
An $n$th order quasilocal charge becomes  
\begin{equation}
 Q^{(n)}[\partial\Sigma] = \int_\Sigma J^{(n)}{}^a d\Sigma_a 
  = \frac{1}{2^{n+2}(D-2n-1)} \oint_{\partial\Sigma} F^{(n)ab}dS_{ab} .
\end{equation}

We note that the potential $2$-form field (\ref{eq:potential_two-form})
seems to be very similar to a part of the Killing-Lovelock potential~\cite{Kastor:2008xb,Kastor:2010gq} to define improved Komar integrals in Lovelock theory.
The $n$th Killing-Lovelock potential for the $n$th order Lovelock term, however, consists of ($n-1$) powers of the Riemann tensor.
On the other hand, in Ref.~\cite{Ozkarsligil:2023avt}, the authors introduced a $2$-form field with the same powers of the Riemann tensor as (\ref{eq:potential_two-form}) for Killing-Yano $2$-forms but not for conformal Killing-Yano $2$-forms. 
In that case, the $2$-form field itself is conserved.

 \section{Applications to known examples}
 \label{sec:examples}

 In this section, we will demonstrate that, for the conventional Kodama vectors, which were heuristically obtained in specific spacetimes, various properties can be reproduced in terms of CKY $2$-forms admitted by those spacetimes. In particular, such spacetimes admit closed CKY $2$-forms, that is, a subclass of CKY $2$-forms. 

  \subsection{Warped-product spacetime}
  
  It is known that warped-product spacetimes with two-dimensional base
  possess the Kodama vector field. 
  We revisit the known results for the
  Kodama vector in terms of CKY $2$-forms (also see Appendix D
  in~\cite{Nozawa:2008rjk}).
  
  We consider that the metric of a $D$-dimensional warped-product spacetime, 
  $\mathcal{B}\times_r \mathcal{F}$, is given by 
  \begin{equation}
   g_{ab}dx^a dx^b = \gamma_{\mu\nu}(y)dy^\mu dy^\nu
    + r(y)^2 \omega_{IJ}(\sigma) d\sigma^I d\sigma^J ,
    \label{eq:warped_product}
  \end{equation}
  where $\gamma_{\mu\nu}$ and $\omega_{IJ}$ denote metrics on the
  two-dimensional base space $\mathcal{B}$ and the $(D-2)$-dimensional
  fiber $\mathcal{F}$, respectively. 
  The positive function $r(y)$ is a warp factor depending only on the
  coordinates on the base space, $\{y^\mu\}$.
  On $\mathcal{F}$ the metric $\omega_{IJ}$ itself becomes a rank-$2$ Killing
  tensor and the associated $(D-2)$-dimensional volume form is a
  Killing-Yano $(D-2)$-form.
  It follows from the lifting theorem in~\cite{Krtous:2015ona} that we can lift
  it to a Killing-Yano $(D-2)$-form on the whole spacetime.
  As a result, we find that this spacetime admits a CCKY $2$-form given by 
  \begin{equation}
   \frac{1}{2} h_{ab} dx^a \wedge dx^b \equiv  
    \frac{r}{2} \, {}^{(\gamma)}\epsilon_{ab} dx^a \wedge dx^b =
    r \sqrt{-\gamma} dy^0 \wedge dy^1
    ,
    \label{eq:CCKY_2-form}
  \end{equation}
  where ${}^{(\gamma)}\epsilon_{ab}$ is the two-dimensional volume form
  associated with the metric $\gamma_{\mu\nu}$.
  Note that this is equivalent to the Hodge dual $f=*h$ being the Killing-Yano ($D-2$)-form.
  
  The associated vector with this CCKY $2$-form yields the Kodama vector as follows: 
  \begin{equation}
   \nabla_a h^{ab} = \frac{1}{\sqrt{-g}} \partial_a 
    \left(\sqrt{-g} r {}^{(\gamma)}\epsilon^{ab}\right)
    = \frac{1}{r^{D-2}\sqrt{-\gamma}\sqrt{\omega}}
    \partial_a 
    \left(r^{D-1}\sqrt{-\gamma}\sqrt{\omega} 
     {}^{(\gamma)}\epsilon^{ab}\right)
    = (D-1) {}^{(\gamma)} \epsilon^{ab} \nabla_a r ,
  \end{equation}
  where the conventional Kodama vector is given by 
  $K^a = - {}^{(\gamma)}\epsilon^{ab} \nabla_b r$.
  In fact, the warp factor $r$ is given by a ``norm'' of the CCKY
  $2$-form $h$ [or the KY ($D-2$)-form $f$] as 
  \begin{equation}
   r^2 = - \frac{1}{2}h_{ab}h^{ab} .
  \end{equation}

  For the Einstein tensor, the components on the two-dimensional base space
  are 
  \begin{equation}
   G_{\mu\nu} = - \frac{D-2}{r} \bar\nabla_\mu \bar\nabla_\nu r + 
    \left[\frac{(D-2)(D-3)}{2r^2} \bar\nabla^\lambda r
     \bar\nabla_\lambda r + \frac{D-2}{r} 
     \bar\nabla_\lambda\bar\nabla^\lambda r
     - \frac{1}{2r^2} {}^{(\omega)}R\right] \gamma_{\mu\nu} ,
  \end{equation}
  where $\bar\nabla_\mu$ denotes the covariant derivative associated with $\gamma_{\mu\nu}$ and ${}^{(\omega)}R$ is the scalar curvature of the ($D-2$)-dimensional metric $\omega_{IJ}$. 
  For a conserved current $G_{ab} K^b$, we have 
  \begin{equation}
   \begin{aligned}
    G_{ab} K^b 
    &= - \frac{1}{r^{D-2}} {}^{(\gamma)}\epsilon_{ab} \nabla^b
    \left[
    (D-2)\frac{r^{D-3}}{2} \nabla^c r \nabla_c r 
     - \frac{r^{D-3}}{2(D-3)} {}^{(\omega)}R
    \right] \\
    &= \frac{1}{r^{D-1}} h_{ab} \nabla^b m ,
   \end{aligned}
  \end{equation}
  where 
  a mass function can be defined by 
  \begin{equation}
   m = \frac{D-2}{2}r^{D-3} 
    \left[ K^a K_a
     + \frac{{}^{(\omega)}R}{(D-2)(D-3)}\right] .
  \end{equation}

  Since $K^a$ is divergence-free, the Kodama vector itself becomes a
  conserved current for the metric tensor $g_{ab}$.
  By definition, a charge associated with this current is given by 
  \begin{equation}
   K_a = - \frac{1}{r} h_{ab}\nabla^b r
    = - \frac{1}{(D-1)r^{D-1}} h_{ab} \nabla^b r^{D-1} .
  \end{equation}
  If we consider the Einstein equation with a cosmological constant
  term, $G_{ab} + \Lambda g_{ab}=T_{ab}$, the Misner-Sharp quasilocal
  mass 
  \begin{equation}
    m_\text{MS} = \frac{D-2}{2}r^{D-3} 
    \left[- \frac{2\Lambda}{(D-1)(D-2)}r^2 + K^a K_a
     + \frac{{}^{(\omega)}R}{(D-2)(D-3)}\right] 
  \end{equation}
  is obtained by combining two conserved charges, including only the
  contribution of matter without a cosmological constant.
  It is built from the CCKY $2$-form and the Ricci scalar on the fiber
  $\mathcal{F}$.

  \subsection{Three-dimensional spacetime}
  
  In three dimensions one can consider that spacetimes are not warped product
  but axisymmetric, such as a rotating spacetime with angular momentum. 
  In this case the Kodama vector can be defined and it provides
  conserved current and charge~\cite{Gundlach:2021six,Kinoshita:2021qsv}.

  Let us suppose $\psi_a$ is a Killing vector satisfying 
  \begin{equation}
   \nabla_a \psi_b + \nabla_b \psi_a = 0 .
  \end{equation}
  The Hodge dual of it provides a CCKY $2$-form given by 
  \begin{equation}
   h_{ab} \equiv \epsilon_{abc} \psi^c .
  \end{equation}
  Note that we can directly confirm 
  \begin{equation}
   \begin{aligned}
    \nabla_c h_{ab} &= \epsilon_{abd} \nabla_c \psi^d \\
    &= g_{ac} K_b - g_{bc} K_a ,
   \end{aligned}
  \end{equation}
  where 
  the associated vector is given by 
  \begin{equation}
   K_a \equiv - \frac{1}{2}\nabla^b h_{ab}
    = - \frac{1}{2}\epsilon_{abc}\nabla^b \psi^c .
  \end{equation}
  This is the extended Kodama vector, which has been introduced in~\cite{Gundlach:2021six,Kinoshita:2021qsv}.
  Note that $\nabla_a \psi_b = \epsilon_{abc}K^c$.
  We have 
  \begin{equation}
   \begin{aligned}
    \nabla_a K_b &= -
    \frac{1}{2}\epsilon_{b}{}^{cd}\nabla_a\nabla_c\psi_d \\
    &= \frac{1}{2}\epsilon_b{}^{cd}R_{cda}{}^e\psi_e
    \\
    &= - 
    \epsilon_{acd} G_b{}^{d}
    \psi^c
    = h_{ac}G_b{}^c ,
   \end{aligned}
  \end{equation}
  which yields $G^{ab}\nabla_aK_b = 0$.
  A straightforward calculation shows 
  \begin{equation}
   \begin{aligned}
    \nabla_a (K^bK_b) &= 2 K^b \nabla_a K_b \\
    &= - 2 \epsilon_{acd}\psi^c G^d{}_b K^b
    = 2 h_{ac}G^{c}{}_b K^b ,
   \end{aligned}
  \end{equation}
  \begin{equation}
   \begin{aligned}
    \nabla_a (\psi^b \psi_b) &= 2 \psi^b \nabla_a \psi_b \\
    &= 2 \epsilon_{abc}\psi^b K^c
    = -2h_{ab}K^b ,
   \end{aligned}
  \end{equation}
  and
  \begin{equation}
   \begin{aligned}
    \nabla_a (\psi^b K_b) 
    &= \psi^b \nabla_a K_b + K_b \nabla_a \psi^b 
    \\
    &= - \epsilon_{acd}\psi^c G_b{}^d \psi^b
    = h_{ac}G^c{}_b\psi^b .
   \end{aligned}
  \end{equation}
  This implies that the above scalar quantities $K^aK_a$, $\psi^a\psi_a$,
  and $\psi^aK_a$ are conserved charges associated with conserved
  currents 
  $G^a{}_b K^b$, $K^a$, and $G^a{}_b \psi^b$, respectively.

  If we assume that $\psi^a$ is an axial Killing vector and the Einstein
  equation $G_{ab} + \Lambda g_{ab} = T_{ab}$ is satisfied, 
  the following scalar functions 
  \begin{equation}
   \begin{aligned}
    m &\equiv \frac{1}{2}( - \Lambda \psi^a\psi_a + K^aK_a) ,\\
    j &\equiv - \psi^a K_a ,
   \end{aligned}
  \end{equation}
  can be interpreted as a Misner-Sharp quasilocal mass and Komar
  angular momentum in three-dimensional axisymmetric spacetimes.

  \subsection{Generalized Misner-Sharp mass in Lovelock gravity}
  
  In this subsection we consider $D$-dimensional warped-product
  spacetime (\ref{eq:warped_product}) again.
  For simplicity, we focus on the cases in which the metric
  $\omega_{IJ}$ on the ($D-2$)-dimensional
  subspace $\mathcal{F}$ is maximally symmetric, i.e., 
  ${}^{(\omega)}R=(D-2)(D-3)k$. 
  The real constant $k$ denotes a curvature scale on the ($D-2$)-dimensional subspace. 
  
  Now, because the whole spacetime is warped product, 
  components of
  $2$-form potential only on the two-dimensional base should contribute
  to the conserved charge by integrating the conserved current for the
  $n$th Lovelock tensor.
  The CCKY $2$-form of Eq.~(\ref{eq:CCKY_2-form}), $h_{ab}$, is proportional to the volume form of the two-dimensional base space.
  We have 
  \begin{equation}
   \begin{aligned}
    F^{(n)}_{ab}h^{ab} &= 
    \delta^{cda_1\cdots a_{2n}}_{abb_1\cdots b_{2n}} 
    h^{ab} h_{cd}
    R_{a_1a_2}{}^{b_1b_2}\cdots R_{a_{2n-1}a_{2n}}{}^{b_{2n-1}b_{2n}} \\
    &= -4r^2 \delta^{I_1 \cdots I_{2n}}_{J_1 \cdots J_{2n}}
    R_{I_1I_2}{}^{J_1J_2} \cdots R_{I_{2n-1}I_{2n}}{}^{J_{2n-1}J_{2n}} \\
    &= -\frac{(D-2)!2^{n+2}}{(D-2n-2)!r^{2n-2}}(k+K^aK_a)^n ,
   \end{aligned}
  \end{equation}
  where ($D-2$)-dimensional 
  components of the Riemann curvature tensor are given by 
  \begin{equation}
   R_{IJ}{}^{KL} = \frac{k + K^aK_a}{r^2} \delta^{KL}_{IJ} .
  \end{equation}
  Note that $\delta^{KL}_{IJ}$ denotes the generalized Kronecker $\delta$ symbol
  on the $(D-2)$ dimensions, and we have used the formulas  
  $\epsilon_{abc_1\cdots c_{D-2}}h^{ab} = -2r^{D-1}{}^{(\omega)}\epsilon_{c_1\cdots c_{D-2}}$ and 
  $\delta^{I_1I_2 \cdots I_{2n-1}I_{2n}}_{J_1J_2\cdots J_{2n-1}J_{2n}}\delta^{J_1J_2}_{I_1I_2}\cdots\delta^{J_{2n-1}J_{2n}}_{I_{2n-1}I_{2n}} = 2^n(D-2)!/(D-2n-2)!$.
  As the result, a conserved current and a quasilocal charge for the $n$th
  Lovelock tensor are 
  \begin{equation}
   G^{(n)a}{}_b K^b = \nabla_b
    \left(\frac{m^{(n)}}{r^{D-1}}h^{ab}\right) ,
  \end{equation}
  where 
  \begin{equation}
   m^{(n)} \equiv 
     \frac{(D-2)!}{2(D-2n-1)!}r^{D-2n-1}(k+K^aK_a)^n .
  \end{equation}
Since, for each order of the Lovelock tensor, each current and each charge
  are conserved, linear combinations of these quantities should be conserved.
  Hence, according to the field equations, they can
  reproduce the generalized Misner-Sharp quasilocal mass in Lovelock gravity, which has been proposed in Refs.~\cite{Maeda:2007uu,Maeda:2011ii}. 

  We note that, when the ($D-2$)-dimensional subspace is described by
  Einstein spaces as well as maximally symmetric spaces, the
  Misner-Sharp quasilocal mass was provided in
  Ref.~\cite{Ohashi:2015xaa}.
  In that case, the quasilocal mass contains the Weyl curvature of the
  ($D-2$)-dimensional Einstein space.
  [More generally, it comprises the sum of every order of Lovelock terms
  for the ($D-2$)-dimensional subspace, as shown in Appendix~\ref{app:A}.]

  \section{discussion}

  In this paper, we have shown that the associated vector of a conformal
  Killing-Yano $2$-form is the origin of the Kodama vector.
  In spacetimes admitting a CKY $2$-form, each order of the Lovelock
  tensors as well as the Einstein tensor contracted with the Kodama
  vector yields a locally conserved current.
  This fact results from purely geometrical properties of CKY forms
  without the field equations in gravitational theories.
  Physical interpretations of the conserved current such as an energy
  current should be provided through the field equations.
  The Kodama vectors that have been known in the literature
  arise from closed CKY $2$-forms. 
  This means that in order to obtain characteristic properties of the Kodama vectors only weaker conditions need to be imposed on spacetimes because closed CKY $2$-forms are contained within CKY $2$-forms. 
  We expect that various arguments based on the Kodama vector can be
  extended to spacetimes admitting CKY $2$-forms as well as closed ones.
  Unfortunately, little is known about general ansatz of nontrivial
  spacetimes admitting a CKY $2$-form such that its associated vector is
  not Killing vector.
  If a spacetime admits a CCKY $2$-form, we can obtain the spacetime
  admitting the CKY $2$-form by conformal transformation.
  Thus, it turns out that conformally warped-product spacetimes have the Kodama vectors. 

  For each order of the Lovelock tensor, including the Einstein tensor
  and metric tensor (i.e., cosmological constant term), each current
  provided by the Kodama vector can be individually conserved. 
  This means there are individual, conserved charges associated with
  each current.
  It is expected that in terms of these charges we can obtain
  thermodynamic relations such as the Smarr formula and the first law~\cite{Maeda:2011ii,Saavedra:2023rfq}.
  In particular, this nature may play a significant role in extracting the
  contribution of a cosmological constant from a definition of energy~\cite{Altas:2018pkl,Petrov:2019roe}. 

  The conserved currents associated with Killing vectors and Kodama
  vectors have a similar structure~\cite{Kastor:2008xb,Kastor:2010gq}
  built from the following quantities: 
  $\mathcal{P}^{(n)}{}_{ab}{}^{cd}\equiv \delta^{cda_1\cdots
  a_{2n}}_{abb_1\cdots b_{2n}} R_{a_1a_2}{}^{b_1b_2}\cdots R_{a_{2n-1}a_{2n}}{}^{b_{2n-1}b_{2n}}$,  
  which are crucial to the Euler-Lagrange equations in Lovelock gravity~\cite{Mukhopadhyay:2006vu,Dadhich:2008df,Padmanabhan:2013xyr}.
  Because a Killing vector $\xi^a$ is divergence-free, we obtain a $2$-form potential $\omega_{ab}$ such that $\xi^a = \nabla_b \omega^{ab}$.
  A part of a Komar-type potential is given by 
  $\mathcal{P}^{(n)}_{abcd}\omega^{cd}$.
  On the other hand, a Kodama vector is provided by a CKY $2$-form $h_{ab}$ as 
  $K^a = -\nabla_b h^{ab}/(D-1)$ and the potential is given by 
  $\mathcal{P}^{(n)}_{abcd} h^{cd}$.
  It is fascinating to explore the relation between these conserved currents.

  A primitive proof of Birkhoff's theorem based on the CKY $2$-form
  can apply to only vacuum with a cosmological constant not but
  electrovac spacetimes, because it relies on the fact that the spacetime
  is described by the Einstein metric.
  However, the condition that the spacetime is described by the Einstein metric is
  only a sufficient condition for the Kodama vector to be a Killing vector.
  The fact that Birkhoff's theorem holds for a wider class of spacetimes even
  in Lovelock gravity~\cite{Zegers:2005vx,Deser:2005gr,Ray:2015ava}
  implies the proof can be improved.
  For example, extending the argument to generalized CKYs or CCKYs with
  torsion~\cite{Kubiznak:2009qi,Houri:2010fr} may be interesting.

 \begin{acknowledgments}
  I would like to thank T.~Houri, M.~Nozawa, and N.~Dadhich for valuable comments.
  This work was supported in part by JSPS KAKENHI Grant No.~JP21H05186. 
 \end{acknowledgments}

\appendix

 \section{Curvature tensors on warped-product spacetimes}
 \label{app:A}

 In this appendix, we summarize useful relations in terms of the
 curvature tensor on the $D$-dimensional warped-product spacetimes,
 $\mathcal{B}\times_r \mathcal{F}$,
 described by the metric (\ref{eq:warped_product}).

 The nonvanishing components of the Riemann tensor are given by 
 \begin{equation}
  \begin{aligned}
   R_{\mu\nu\alpha\beta} &= {}^{(\gamma)}R
   \gamma_{\mu[\alpha}\gamma_{\beta]\nu} ,\\
   R_{\mu I \nu J} &= - \omega_{IJ} r \bar\nabla_\mu \bar\nabla_\nu r ,\\
   R_{IJKL} &= r^2 \left[{}^{(\omega)}R_{IJKL}
   + 2 K^aK_a \omega_{I[K}\omega_{L]J}
   \right] ,
   \end{aligned}
 \end{equation}
 where ${}^{(\omega)}R_{IJKL}$ denotes the Riemann tensor with respect
 to the metric $\omega_{IJ}$ on the fiber $\mathcal{F}$, and ${}^{(\gamma)}R$ and
 $\bar\nabla_\mu$ denote, respectively, the Ricci scalar and the covariant derivative
 with respect to the metric $\gamma_{\mu\nu}$ on the base $\mathcal{B}$.

 The nonvanishing components of the Ricci tensor are 
 \begin{equation}
  \begin{aligned}
   R_{\mu\nu} &= \frac{{}^{(\gamma)}R}{2}\gamma_{\mu\nu}
   - \frac{D-2}{r} \bar\nabla_\mu \bar\nabla_\nu r ,\\
   R_{IJ} &= 
   {}^{(\omega)}R_{IJ}
   + \left[(D-3) K^aK_a - r \bar\nabla^\mu \bar\nabla_\mu r
   \right]\omega_{IJ} ,
  \end{aligned}
 \end{equation}
 and the Ricci scalar is 
 \begin{equation}
  R = {}^{(\gamma)}R
   - \frac{2(D-2)}{r} \bar\nabla^\mu \bar\nabla_\mu r
   + \frac{1}{r^2}
   \left[
    {}^{(\omega)}R
    + (D-2)(D-3) K^aK_a 
   \right] .
 \end{equation}

 The contraction of $n$ powers of the Riemann tensor with the generalized Kronecker $\delta$ on ($D-2$)-space is given by  
 \begin{equation}
  \begin{aligned}
    \delta^{I_1 \cdots I_{2n}}_{J_1 \cdots J_{2n}}
    R_{I_1I_2}{}^{J_1J_2} \cdots R_{I_{2n-1}I_{2n}}{}^{J_{2n-1}J_{2n}} 
   &= \frac{1}{r^{2n}}
   \delta^{I_1 \cdots I_{2n}}_{J_1 \cdots J_{2n}}
   \left[{}^{(\omega)}R_{I_1I_2}{}^{J_1J_2}
   + K^aK_a \delta_{I_1I_2}^{J_1J_2}
   \right] \cdots \\
   &= \frac{2^n}{r^{2n}} \sum_{l=0}^{n}
   {}_n C_l \frac{(D-2-2n+2l)!}{(D-2-2n)!}(K^aK_a)^l
   \mathcal{R}^{(n-l)}_\omega ,
  \end{aligned}
  \label{eq:dRR}
 \end{equation}
 where 
 \begin{equation}
  \mathcal{R}_\omega^{(k)} \equiv \frac{1}{2^k} 
   \delta^{I_1 \cdots I_{2k}}_{J_1 \cdots J_{2k}}
    {}^{(\omega)}R_{I_1I_2}{}^{J_1J_2} \cdots 
    {}^{(\omega)}R_{I_{2k-1}I_{2k}}{}^{J_{2k-1}J_{2k}} .
    \label{eq:Lovelock_omega}
 \end{equation}
 Note that we have used 
 \begin{equation}
  \delta^{I_1 \cdots I_{2n}}_{J_1 \cdots J_{2n}}
   \delta^{I_1 I_{2}}_{J_1 J_{2}}
   \cdots \delta^{I_{2l-1} I_{2l}}_{J_{2l-1} J_{2l}}
   =
   \frac{2^l (D-2-2n+2l)!}{(D-2-2n)!}
   \delta^{I_1 \cdots I_{2(n-l)}}_{J_1 \cdots J_{2(n-l)}}
   \qquad (n\ge l) .
 \end{equation}
 If $\omega_{IJ}$ on the fiber $\mathcal{F}$ is a ($D-2$)-dimensional 
 Einstein metric, i.e, 
 ${}^{(\omega)}R_{IJ} = k(D-3) \omega_{IJ}$,
 then we have 
 ${}^{(\omega)}R_{IJ}{}^{KL} = {}^{(\omega)}W_{IJ}{}^{KL} + k \delta_{IJ}^{KL}$,
 where ${}^{(\omega)}W_{IJKL}$ is the Weyl tensor with respect to $\omega_{IJ}$.
 Equation~(\ref{eq:dRR}) reduces to  
 \begin{equation}
  \delta^{I_1 \cdots I_{2n}}_{J_1 \cdots J_{2n}}
   R_{I_1I_2}{}^{J_1J_2} \cdots R_{I_{2n-1}I_{2n}}{}^{J_{2n-1}J_{2n}} 
   = \frac{2^n}{r^{2n}} \sum_{l=0}^{n}
   {}_n C_l \frac{(D-2-2n+2l)!}{(D-2-2n)!}(k + K^aK_a)^l
   \mathcal{W}^{(n-l)}_\omega ,
 \end{equation}
 where $\mathcal{W}_\omega^{(k)}$ has been obtained by replacing 
 the Riemann tensor ${}^{(\omega)}R_{IJ}{}^{KL}$ with 
 the Weyl tensor ${}^{(\omega)}W_{IJ}{}^{KL}$ in Eq.~(\ref{eq:Lovelock_omega}).

 \section{Formulas for curvature polynomials}

 In this appendix, we summarize basic properties of curvature polynomials in $D$-dimensions
 (see, for example, Refs.~\cite{Padmanabhan:2013xyr,Mukhopadhyay:2006vu,Dadhich:2008df}).

 The $n$th order Lovelock scalar and Lovelock-Ricci tensor, 
 which are respectively analogous to the Ricci scalar and the Ricci tensor for $n=1$, are 
 \begin{align}
  \mathcal{R}^{(n)} 
  &\equiv \frac{1}{2^n} \delta^{a_1a_2\cdots a_{2n}}_{b_1b_2\cdots b_{2n}}
   R_{a_1a_2}{}^{b_1b_2} \cdots R_{a_{2n-1}a_{2n}}{}^{b_{2n-1}b_{2n}} ,\\
  \mathcal{R}^{(n)a}{}_b &\equiv \frac{n}{2^n} 
  \delta^{a_1a_2\cdots a_{2n}}_{b b_2\cdots b_{2n}}
   R_{a_1a_2}{}^{a b_2} \cdots R_{a_{2n-1}a_{2n}}{}^{b_{2n-1}b_{2n}} .
 \end{align}
 The $n$th order Lovelock tensor, which is the analog of the Einstein tensor, is given by 
 \begin{equation}
  \begin{aligned}
     G^{(n)a}{}_b &\equiv - \frac{1}{2^{n+1}}\delta^{aa_1a_2\cdots a_{2n}}_{bb_1b_2\cdots b_{2n}}
   R_{a_1a_2}{}^{b_1b_2} \cdots R_{a_{2n-1}a_{2n}}{}^{b_{2n-1}b_{2n}} \\
   &= \mathcal{R}^{(n)a}{}_b - \frac{1}{2} \mathcal{R}^{(n)} g^a{}_b ,
  \end{aligned}
 \end{equation}
 where the last equality is easily verified by using the following formula:  
 \begin{equation}
  \delta^{a a_1a_2\cdots a_{2n}}_{b b_1b_2\cdots b_{2n}}
   = g^a_b \delta^{a_1a_2\cdots a_{2n}}_{b_1b_2\cdots b_{2n}}
   - \sum^{2n}_{k=1} g^a_{b_k}
   \delta^{a_1a_2\cdots a_k \cdots a_{2n}}_{b_1b_2 \cdots b \cdots b_{2n}}.
 \end{equation}
 
 An $n$th order rank-$4$ tensor that consists of $n$ powers of the Riemann tensor is given by 
 \begin{equation}
  \mathcal{P}^{(n)}{}_{ab}{}^{cd} \equiv 
    \delta^{cda_1a_2\cdots a_{2n}}_{abb_1b_2\cdots b_{2n}}
   R_{a_1a_2}{}^{b_1b_2} \cdots R_{a_{2n-1}a_{2n}}{}^{b_{2n-1}b_{2n}} .
 \end{equation}
 Its indices have the same properties as those of the Riemann tensor,
 \begin{equation}
  \mathcal{P}^{(n)}{}_{abcd} = - \mathcal{P}^{(n)}{}_{bacd}
   = - \mathcal{P}^{(n)}{}_{abdc}, \quad
   \mathcal{P}^{(n)}{}_{abcd} =  \mathcal{P}^{(n)}{}_{cdab} ,\quad
   \mathcal{P}^{(n)}{}_{[abc]d} = 0 ,
 \end{equation}
 and, in addition, it is divergence-free for each index, 
 \begin{equation}
  \nabla^a\mathcal{P}^{(n)}{}_{abcd} = 0 .
 \end{equation}
 This tensor has various useful properties as follows.
 The contraction yields 
 \begin{equation}
  \mathcal{P}^{(n)}{}_{ac}{}^{bc} = - 2^{n+1}(D-2n-1) G^{(n)}{}_a{}^b . 
 \end{equation}
 Furthermore, we have 
 \begin{equation}
  \mathcal{R}^{(n)} = \frac{1}{2^n} \mathcal{P}^{(n-1)}{}_{abcd} R^{abcd} ,
\quad
\mathcal{R}^{(n)a}{}_b = \frac{n}{2^n} \mathcal{P}^{(n-1)}{}_{bcde} R^{acde}
.
 \end{equation}

\bibliography{references}

\end{document}